\documentstyle[amssymb,prl,multicol,psfig,aps]{revtex}

\begin{document}
\title{Spin-orbit coupling effects on quantum transport in lateral semiconductor
dots}
\author{I.L. Aleiner$^{1,2}$ and Vladimir I. Fal'ko$^{1}$}
\address{$^{1}$ Physics Department, Lancaster University, LA1 4YB, Lancaster, UK\\
$^{2}$ Physics and Astronomy Department, SUNY at Stony Brook, Stony Brook,\\
NY 11794, USA}
\date{\today}
\maketitle

\begin{abstract}
The effects of interplay between spin-orbit coupling and Zeeman splitting on
weak localisation and universal conductance fluctuations in lateral
semiconductor quantum dots are analysed: All possible symmetry classes of
corresponding random matrix theories are listed and crossovers between them
achievable by sweeping magnetic field and changing the dot parameters are
described. We also suggest  experiments to measure the 
spin-orbit coupling constants.
\end{abstract}

\draft

\begin{multicols}{2}
The effects of spin-orbit (SO) coupling on
transport phenomena in chaotic quantum dots \cite{MarcusSpin,Halperin} recently
attracted attention.
Motivated by a puzzling modification of the variance of the mesoscopic
conductance fluctuations with applied in-plane magnetic field, Halperin
{\it  et. al. }\cite{Halperin} suggested that the specific form of the spin-orbit
interaction in a 2D electron gas based on semiconductor heterostructures may
be responsible for a series of crossovers not considered in the existing
literature \cite{Beenakker}. It has also been noticed 
\cite{Nazarov,Halperin} that spin relaxation in a quantum dot may be facilitated
by the Zeeman field.

The goal of this paper is two-fold: (i) we provide a complete quantitative
theory for the interplay between spin-orbit coupling effects and Zeeman
splitting in zero-dimensional semiconducting systems, identify all the
possible symmetries, and describe all the physically achievable crossovers;
and (ii) we show that the SO coupling effects depend on the magnetic field
orientation in anisotropic dots and discuss possible experiments enabling
one to measure directly the ratio between two independent SO constants.

The single-particle Hamiltonian of the system, $H=H_{0}+u(\vec{r})$ is the
sum of the free-electron dispersion term and a potential, $u(\vec{r})$,
consisting of a confining potential and a random potential of impurities.
The free-electron term includes spin-orbit coupling, as a combination of a
Rashba term and a crystalline anisotropy term (specified for (001) plane of
GaAs), and Zeeman splitting energy due to the in-plane magnetic field \cite
{Remark-perpfield}, $\vec{B}=\vec{l}B$,

\[
H_{0}=\frac{p^{2}}{2m}+\frac{\alpha }{m}[\vec{p}{\bf \times }\vec{n}_{z}]%
\frac{\vec{\sigma}}{2}+\frac{\varrho }{m}(p_{x}\frac{\sigma _{x}}{2}-p_{y}%
\frac{\sigma _{y}}{2})+\vec{l}\frac{\vec{\sigma}}{2}\epsilon _{{\rm Z}}, 
\]
where $\vec{p}=\vec{P}-e\vec{A}$ is the kinetic momentum, with $\vec{P}$
being the canonical momentum and $\vec{A}$ being the vector potential
describing the orbital effect of the magnetic field. Since (001) plane of
GaAs has the symmetry of a square without inversion centre, $C_{2v}$, we
choose the coordinate system ($x_{1},x_{2}$) with axes along
crystallographic directions $\vec{e}_{1}=[110]$ and $\vec{e}_{2}=[1\bar{1}0]$
and rewrite $H_{0}$ as 
\begin{equation}
H_{0}=\frac{1}{2m}\left[ \left( p_{1}-\frac{\sigma _{2}}{2\lambda _{1}}%
\right) ^{2}+\left( p_{2}+\frac{\sigma _{1}}{2\lambda _{2}}\right) ^{2}%
\right] +\vec{l}\frac{\vec{\sigma}}{2}\epsilon _{{\rm Z}},  \label{H0}
\end{equation}
where $\lambda _{1,2}^{-1}=\alpha \pm \varrho $ characterise the length
scale associated with the strength of the spin-orbit coupling for electrons
moving along principal crystallographic directions [$\sigma _{1,2,3}$ are
Pauli matrices, $\sigma _{2}=-\sigma _{2}^{T}$, $\sigma _{1,3}=\sigma
_{1,3}^{T}$].

The Hamiltonian $H=H_{0}+u(\vec{r})$ describes the electron motion in a
lateral semiconductor dot coupled to metallic leads via two contacts, $l$
and $r$, each with $N_{l,r}\gtrsim 1$ open orbital channels. Below we focus
on the universal 0D description applicable if \cite{Largesample} 
\begin{equation}
\gamma ,\,\epsilon _{{\rm Z}}\ll E_{{\rm T}};\;L_{1,2}\ll \lambda _{1,2}.
\label{escape}
\end{equation}
Here, $\gamma =(N_{l}+N_{r})\Delta /2\pi \hbar $ stands for the escape rate
into the leads, $\Delta =2\pi \hbar ^{2}/m{\cal A}$\ is the mean level
spacing in a dot with area ${\cal A}\sim L_{1}L_{2}$, and $E_{{\rm T}}$ is
the conventional Thouless energy.

Our purpose now is to identify Hamiltonian (\ref{H0}) with an appropriate
random matrix ensemble. Doing it directly, however, is not convenient. The
reason for this is that on shell matrix elements of the velocity vanish due
to the gauge invariance (the importance of this fact for SO interaction in
quantum dots was first noticed in Ref. \cite{Halperin}). It means that if
the spin remained fixed during the motion of the electron, the effect of SO
coupling would be just a homogeneous shift of the momentum space which could
not change observables. To get rid of such terms fixed by gauge invariance,
we perform the unitary transformation of the Hamiltonian as $H\rightarrow 
\tilde{H}=U^{\dagger }HU$ \ with

\begin{equation}
U=\exp \left( \frac{ix_{1}\sigma _{2}}{2\lambda _{1}}-\frac{ix_{2}\sigma _{1}%
}{2\lambda _{2}}\right)  \label{U}
\end{equation}

Using the condition $L_{1,2}/\lambda _{1,2}\ll 1$, we expand $\tilde{H}$ up
to the second order in the coordinates and obtain

\begin{eqnarray}
\tilde{H} &=&\frac{1}{2m}\left( \vec{p}-e\vec{A}-\vec{a}_{\bot }\frac{\sigma
_{z}}{2}-\vec{a}_{\Vert }\right) ^{2}+h^{(0)}+h^{(1)}+u(\vec{r})  \nonumber
\\
\vec{A} &=&B_{z}[\vec{r}{\bf \times }\vec{n}_{z}]/2c;\hspace{0.12in}\vec{a}%
_{\bot }=[\vec{r}{\bf \times }\vec{n}_{z}]/2\lambda _{1}\lambda _{2};\hspace{%
0.12in}  \label{aperp} \\
\vec{a}_{\Vert } &=&\frac{1}{6}\frac{[\vec{r}{\bf \times }\vec{n}_{z}]}{%
\lambda _{1}\lambda _{2}}\left( \frac{x_{1}\sigma _{1}}{\lambda _{1}}+\frac{%
x_{2}\sigma _{2}}{\lambda _{2}}\right)  \label{apar} \\
h^{(0)} &=&\;\epsilon _{{\rm Z}}\vec{l}\frac{\vec{\sigma}}{2};\hspace{0.12in}%
h^{(1)}=-\sigma _{z}\frac{\epsilon _{Z}}{2}\left( \frac{l_{1}x_{1}}{2\lambda
_{1}}+\frac{l_{2}x_{2}}{2\lambda _{2}}\right) .  \label{zeeman}
\end{eqnarray}
Equation (\ref{aperp}) indicates that the effects of SO coupling in the
leading order at $\epsilon _{{\rm Z}}=0$, and of the orbital magnetic field
are somewhat similar. This similarity is not a coincidence --- in the
leading order, the direction of the spin follows the motion of the electron:
for an electron moving along a closed path, its spin spans the closed path
too. Due to the motion in spin space, an electron accumulates extra Berry
phase equal to the solid angle spanned by the spin. At weak SO coupling,
this area is proportional to the geometrical area encircled by the electron
path in the coordinate space resulting in an effect similar to that of
Aharonov-Bohm flux. This analogy may be put on a quantitative level by
noticing that the two energy scales characterizing both effects

\begin{eqnarray}
\tau _{B}^{-1} &=&\frac{4\pi B_{z}^{2}}{\Delta }\langle |M_{\alpha \beta
}|^{2}\rangle =\kappa E_{{\rm T}}\left( \frac{2eB_{z}{\cal A}}{c\hbar }%
\right) ^{2};  \label{A} \\
\epsilon _{\bot }^{{\rm so}} &=&\kappa E_{{\rm T}}\left( {\cal A}/\lambda
_{1}\lambda _{2}\right) ^{2}
\end{eqnarray}
have the same dependence on the shape and the disorder in the sample. Here, $%
\kappa $ is the coefficient dependent on the geometry and ${\cal A}$ is the
area of the dot. Random quantities $M_{\alpha \beta }$ are the non-diagonal
matrix elements of the magnetic moment of the electron in the dot.

Term (\ref{apar}) is higher order in the SO coupling constant. However, it
has a different symmetry from $\vec{a}_{\bot }$, therefore, its retention is
legitimate. Its physical significance is to provide the
spin-flips and, thus,  the complete spin relaxation, in
contrast to $\vec{a}_{\bot }$ which preserves correlations between spin up
and spin down states. Quantitatively, the effect of $\vec{a}_{\Vert }$ is
characterized by the scale

\begin{equation}
\epsilon _{\Vert }^{{\rm so}}\sim \left[ \left( L_{1}/\lambda _{1}\right)
^{2}+\left( L_{2}/\lambda _{2}\right) ^{2}\right] \epsilon _{\bot }^{{\rm so}%
}\ll \epsilon _{\bot }^{{\rm so}}.
\end{equation}

The effect of the in-plane magnetic field is described by Eq. (\ref{zeeman}). 
It includes the homogenous Zeeman splitting $h^{(0)}$and the combined
effect of the SO interaction and Zeeman splitting described by $h^{(1)}$.
The latter can be envisaged as a deflection of the effective magnetic field
from the direction given by external $\vec{B}$, and it results in the spin
relaxation associated with the energy scale 
\begin{equation}
\epsilon _{\bot }^{{\rm Z}}=\frac{\epsilon _{Z}^{2}}{2\Delta }\sum_{i,j=1,2}%
\frac{l_{i}}{\lambda _{i}}\frac{l_{j}}{\lambda _{j}}\;\Xi _{ij},\;\Xi
_{ij}=\pi \langle x_{i}^{\alpha \beta }x_{j}^{\beta \alpha }\rangle ,
\label{BB}
\end{equation}
where $x_{1,2}^{\alpha \beta }$ are the non-diagonal matrix elements of the
dipole moment of the electron in the dot. Quantity$\ \Xi _{ij}$ depends on
the geometry and on the disorder in the dot and may be estimated as $\Xi
\simeq $ $\Delta L^{2}/E_{{\rm T}}$, so that $\epsilon _{\bot }^{{\rm Z}}\ll
\epsilon _{{\rm Z}}$. A similar energy scale has appeared in recent
publications \cite{Nazarov,Halperin}, however, the symmetry of the
corresponding term $h^{(1)}$ was not indentified.

\end{multicols}
\widetext

\begin{table}[tbp] \centering
\caption{Symmetries of the system in the absense of orbital magnetic field 
effect, $\protect\tau_B \gg \tau_{esc}$.\label{Tab1}}

\begin{tabular}{|c|c|c|c|c|c|c|c|c|}
& Zeeman & Spin-orbit & Additional symmetry of$\;\tilde{H}=\tilde{H}%
^{\dagger }$ & Symm. group & $\beta $ & $\Sigma $ & $s$\  & Applicability \\ 
\hline
1 & $h^{(0,1)}=0$ & $\vec{a}_{\bot ,\Vert }=0$ & $\tilde{H}^{T}=\tilde{H},\;%
\left[ \tilde{H},\sigma _{1,2,3}\right] =0$ & $\frac{{\rm O(N)\otimes O(N)}}{%
{\rm O(N)}}$ & 1 & 1 & 2 & $\epsilon _{{\rm Z}},\;\epsilon _{\bot }^{{\rm so}%
}\ll \gamma $ \\ \hline
2 & $h^{(0,1)}=0$ & $
\begin{array}{c}
\vec{a}_{\bot }\neq 0 \\ 
\vec{a}_{\Vert }=0
\end{array}
$ & $\sigma _{2}\tilde{H}^{T}\sigma _{2}=\tilde{H},\;\left[ \tilde{H},\sigma
_{3}\right] =0$ & $\frac{{\rm U(N)\otimes U(N)}}{{\rm U(N)}}$ & 2 & 1 & 2 & $%
\frac{\epsilon _{{\rm Z}}^{2}}{\epsilon _{\bot }^{{\rm so}}},\;\epsilon
_{\Vert }^{{\rm so}}\ll \gamma \ll \epsilon _{\bot }^{{\rm so}}$ \\ \hline
3 & $h^{(0,1)}=0$ & $\vec{a}_{\bot ,\Vert }\neq 0$ & $\sigma _{2}\tilde{H}%
^{T}\sigma _{2}=\tilde{H}$ & {\rm Sp(2N)} & 4 & 1 & 2 & $\frac{\epsilon _{%
{\rm Z}}^{2}}{\epsilon _{\bot }^{{\rm so}}}\ll \gamma \ll \epsilon _{\Vert
}^{{\rm so}}$ \\ \hline
4 & $
\begin{array}{c}
h^{(0)}\neq 0 \\ 
h^{(1)}=0
\end{array}
$ & $\vec{a}_{\bot ,\Vert }=0$ & $\tilde{H}^{T}=\tilde{H},\;\left[ \tilde{H},%
\vec{B}\vec{\sigma}\right] =0$ & ${\rm O(N)\otimes O(N)}$ & 1 & 1 & 1 & $%
\epsilon _{\bot }^{{\rm Z}},\;\epsilon _{\bot }^{{\rm so}}\ll \gamma \ll
\epsilon _{{\rm Z}}\;$ \\ \hline
5 & $
\begin{array}{c}
h^{(0)}\neq 0 \\ 
h^{(1)}=0
\end{array}
$ & $
\begin{array}{c}
\vec{a}_{\bot }\neq 0 \\ 
\vec{a}_{\Vert }=0
\end{array}
$ & $\sigma _{1}\tilde{H}^{T}\sigma _{1}=\tilde{H}$ & ${\rm O(2N)}$ & 1 & 2
& 1 & $\epsilon _{\bot }^{{\rm Z}},\;\epsilon _{\Vert }^{{\rm so}}\ll \gamma
\ll \epsilon _{\bot }^{{\rm so}},\frac{\epsilon _{{\rm Z}}^{2}}{\epsilon
_{\bot }^{{\rm so}}}$ \\ \hline
6 & $
\begin{array}{c}
h^{(0)}\neq 0 \\ 
h^{(1)}=0
\end{array}
$ & $
\begin{array}{c}
\vec{a}_{\bot }\neq 0 \\ 
\vec{a}_{\Vert }\neq 0
\end{array}
$ & none & ${\rm U(2N)}$ & 2 & 2 & 1 & $\epsilon _{\bot }^{{\rm Z}}\ll
\gamma \ll \epsilon _{\Vert }^{{\rm so}},\;\frac{\epsilon _{{\rm Z}}^{2}}{%
\epsilon _{\bot }^{{\rm so}}}$ \\ \hline
7 & $h^{(0,1)}\neq 0$ & $\vec{a}_{\bot ,\Vert }=0$ & $\sigma _{\bot }\sigma
_{2}\tilde{H}^{T}\sigma _{2}\sigma _{\bot }=\tilde{H};\sigma _{\bot }=\vec{%
\sigma}(\vec{l}_{z}\times \vec{l})$ & ${\rm O(2N)}$ & 1 & 2 & 1 & $\epsilon
_{\bot }^{{\rm so}}\ll \gamma \ll \epsilon _{\bot }^{{\rm Z}}$ \\ \hline
8 & $h^{(0,1)}\neq 0$ & $\vec{a}_{\bot }\neq 0$ & none & ${\rm U(2N)}$ & 2 & 
2 & 1 & $\gamma \ll \epsilon _{\bot }^{{\rm so}},\;\epsilon _{\bot }^{{\rm Z}%
}$ \\ 
\end{tabular}
\end{table}

\begin{multicols}{2}

Having derived a Hamiltonian free of the gauge invariance constraints on the
values of its matrix elements, we identify the symmetries of all relavant
limits. The results are summarized in Tables \ref{Tab1} and \ref{Tab2}
depending on the orbital effect of the magnetic field. In these tables, the
conventional parameter $\beta $ describes time-reversal symmetry of the
orbital motion, $s$ is the Kramers degeneracy parameter, and $\Sigma $ is an
additional parameter characterising  the mixing of states with different
spins for strong Zeeman splitting. Parameters $\beta$, $\Sigma$, and $s$
completely characterize the statistical properties of the transport through
the system as well as spectral correlations of the isolated dot. The
straightforward generalization of the known results, \cite{Beenakker}, gives
the following description of the two-terminal conductance of the dot
(measured in units of $\frac{e^{2}}{2\pi \hbar }$) connected to the leads by
reflectionless contacts with $N_{l}$ and $N_{r}$ orbital channels: 
\begin{equation}
\langle g\rangle =\frac{2\Sigma N_{l}N_{r}}{(N_{l}+N_{r})\Sigma +(\frac{2}{%
\beta }-1)},  \label{conductance}
\end{equation}

\end{multicols}
\widetext

\begin{equation}
\langle \left( \delta g\right) ^{2}\rangle =\left( \frac{s}{\beta \Sigma }%
\right) \frac{\Sigma ^{2}N_{l}N_{r}\left[ \Sigma N_{l}+(\frac{2}{\beta }-1)%
\right] \left[ \Sigma N_{r}+(\frac{2}{\beta }-1)\right] }{\left[
(N_{l}+N_{r})\Sigma +(\frac{2}{\beta }-1)\right] ^{2}\left[
(N_{l}+N_{r})\Sigma +(\frac{4}{\beta }-1)\right] \left[ (N_{l}+N_{r})\Sigma
+(\frac{2}{\beta }-2)\right] }.  \label{variance}
\end{equation}
\begin{multicols}{2}
The parametric dependences of the transport coefficients in Eqs. (\ref
{conductance},\ref{variance}) can be envisaged as a sequence of crossovers
shown below as a function of Zeeman splitting energy and escape rate
(increasing along the horizontal and vertical axes, respectively):

\begin{equation}
\begin{array}{ccccc}
1\,[1u] & \rightarrow & 4\,[4u] & \rightarrow & 7\,[6u] \\ 
\uparrow &  & \uparrow &  & \uparrow \\ 
2\,[2u] & \rightarrow & 5\,[5u] & \rightarrow & 8\,[6u] \\ 
\uparrow &  & \uparrow & \nearrow &  \\ 
3\,[3u] & \rightarrow & 6\,[6u] &  & 
\end{array}
{\cal \;} 
\begin{array}{ccc}
\gamma &  &  \\ 
\Uparrow &  &  \\ 
& \Longrightarrow & \epsilon _{{\rm Z}}
\end{array}
\label{sketch}
\end{equation}

For a large number of channels, $1\ll N_{l}+N_{r}\ll E_{{\rm T}}/\Delta $,
the result of Eqs. (\ref{conductance},\ref{variance}) can be simplified: 
\begin{equation}
g_{{\rm wl}}=-\frac{1-\frac{1}{2}\beta }{\beta \Sigma }\xi ;\;\;\langle
\delta g^{2}\rangle =\frac{s\xi ^{2}}{16\beta \Sigma },\;\xi =\frac{%
4N_{l}N_{r}}{(N_{l}+N_{r})^{2}}  \label{LargeN}
\end{equation}
and the crossover in WL behavior between various symmetry classes in the
Tables \ref{Tab1},\ref{Tab2} sketched in (\ref{sketch}) can be described
quantitatively using a diagrammatic perturbation theory \cite{Hikami}, as we
outline below for the weak localisation (WL) correction.

\end{multicols}
\widetext

\begin{table}[tbp] \centering

\caption{Symmetries of the system in the presense of orbital magnetic field 
effect, $\protect \tau_{B} \ll \tau_{esc}$.\label{Tab2}}

\begin{tabular}{|c|c|c|c|c|c|c|c|c|}
& Zeeman & spin-orbit & additional\ symmetry of$\;\tilde{H}=\tilde{H}%
^{\dagger }$ & symmetry group & $\beta $ & $\Sigma $ & $s$\  & applicability
intervals \\ \hline
1u & $h^{(0,1)}=0$ & $\vec{a}_{\bot ,\Vert }=0$ & $\;\left[ \tilde{H},\sigma
_{1,2,3}\right] =0$ & $\frac{{\rm U(N)\otimes U(N)}}{{\rm U(N)}}$ & 2 & 1 & 2
& $\epsilon _{{\rm Z}},\;\epsilon _{\bot }^{{\rm so}}\ll \gamma $ \\ \hline
2u & $h^{(0,1)}=0$ & $
\begin{array}{c}
\vec{a}_{\bot }\neq 0 \\ 
\vec{a}_{\Vert }=0
\end{array}
$ & $\;\left[ \tilde{H},\sigma _{3}\right] =0$ & ${\rm U(N)\otimes U(N)}$ & 2
& 1 & 1 & $\frac{\epsilon _{{\rm Z}}^{2}}{\epsilon _{\bot }^{{\rm so}}}%
,\;\epsilon _{\Vert }^{{\rm so}}\ll \gamma \ll \epsilon _{\bot }^{{\rm so}}$
\\ \hline
3u & $h^{(0,1)}=0$ & $\vec{a}_{\bot ,\Vert }\neq 0$ & none & {\rm U(2N)} & 2
& 2 & 1 & $\frac{\epsilon _{{\rm Z}}^{2}}{\epsilon _{\bot }^{{\rm so}}}\ll
\gamma \ll \epsilon _{\Vert }^{{\rm so}}$ \\ \hline
4u & $
\begin{array}{c}
h^{(0)}\neq 0 \\ 
h^{(1)}=0
\end{array}
$ & $\vec{a}_{\bot ,\Vert }=0$ & $\;\left[ \tilde{H},\vec{B}\cdot \vec{\sigma%
}\right] =0$ & ${\rm U(N)\otimes U(N)}$ & 2 & 1 & 1 & $\epsilon _{\bot }^{%
{\rm Z}},\;\epsilon _{\bot }^{{\rm so}}\ll \gamma \ll \epsilon _{{\rm Z}}\;$
\\ \hline
5u & $h^{(0)}\neq 0$ & $\vec{a}_{\bot }\neq 0$ & none & ${\rm U(2N)}$ & 2 & 2
& 1 & $\gamma \ll \epsilon _{\bot }^{{\rm so}},\;\frac{\epsilon _{{\rm Z}%
}^{2}}{\epsilon _{\bot }^{{\rm so}}}$ \\ \hline
6u & $h^{(0,1)}\neq 0$ & $\vec{a}_{\bot ,\Vert }=0$ & none & ${\rm U(2N)}$ & 
2 & 2 & 1 & $\epsilon _{\bot }^{{\rm so}}\ll \gamma \ll \epsilon _{\bot }^{%
{\rm Z}}$ \\ 
\end{tabular}
\end{table}

\begin{multicols}{2}

For a spin-$\frac{1}{2}$ particle in a quantum dot with ballistic adiabatic
contacts, the WL correction to the conductance can be related to the
lowest-lying modes of Cooperons in a singlet $(L=0)$, $P_{{\rm C}}^{00}=%
\frac{1}{2}{\rm tr}\langle \sigma _{2}\hat{G}_{R}^{T}(\varepsilon )\sigma
_{2}\hat{G}_{A}(\varepsilon -\omega )\rangle $, and three triplet ($L=1,2,3$%
) channels,    $P_{{\rm C}}^{LM}=\frac{1}{2}{\rm tr}\langle \sigma
_{L}\sigma _{2}\hat{G}_{R}^{T}(\varepsilon )\sigma _{2}\sigma _{M}\hat{G}%
_{A}(\varepsilon -\omega )\rangle $, as$\ g_{{\rm wl}}\propto P_{{\rm C}%
}^{00}-\sum_{M=1,2,3}P_{{\rm C}}^{MM}$ [Ref. \cite{Hikami}]. In the absence
of SO coupling and Zeeman splitting, Cooperons $P_{{\rm C}}^{LM}$ split into
completely independent channels: one singlet and three triplet, $\hat{P}=%
\hat{\delta}P$, where $\hat{\delta}\equiv \delta ^{LM}$ and $P$ obeys the
diffusion equation. The SO coupling and Zeeman splitting mix up various
components \cite{LyandaGeller} and split their spectra, which modifies the
diffusion equation into the matrix equations $\hat{\Pi}\hat{P}(\vec{X},\vec{X%
}^{\prime })=\hat{\delta}\cdot \delta (\vec{X},\vec{X}^{\prime })$, 
\begin{eqnarray}
\hat{\Pi} &=&\gamma \hat{\delta}+i\epsilon _{{\rm Z}}\;\hat{\eta}-D(\hat{%
\delta}\partial _{X_{1}}+i\hat{\delta}2A_{1}-i\hat{S}_{2}\lambda
_{1}^{-1})^{2}  \label{Coop} \\
&&-D(\hat{\delta}\partial _{X_{2}}+i\hat{\delta}2A_{2}+i\hat{S}_{1}\lambda
_{2}^{-1})^{2},  \nonumber
\end{eqnarray}
where $\hat{S}_{K}^{LM}=-i\varepsilon ^{KLM}$ are spin-1 operators ($K=1,2,3$%
), $\varepsilon ^{KLM}$ is the antisymmetric tensor ($K,L,M=1,2,3$), and $%
\eta ^{LM}=l_{L}\delta _{0M}+\delta _{0L}l_{M}$ indicating 
that coherence between oppositely polarized electrons is lost on the time
scale of $\epsilon _{{\rm Z}}^{-1}$, and $D$ is the classical diffusion
coefficient. Equation (\ref{Coop}) is supplemented with the
the boundary condition at the edge of the dot characterised by the normal
direction $\vec{n}_{\Vert }=(n_{1},n_{2})$, \ 
\begin{equation}
\left[ \vec{n}_{\Vert }\cdot \hat{\delta}(\nabla +i2\vec{A})-in_{1}\hat{S}%
_{2}\lambda _{1}^{-1}+in_{2}\hat{S}_{1}\lambda _{2}^{-1}\right] \hat{P}=0.
\label{boundary}
\end{equation}

The correspondence between random matrix theory description of a disordered
system and diagrams is usually transparent in the zero-dimensional (0D)
approximation in the diffusion problem, $E_{{\rm T}}\rightarrow \infty $,
when the lowest modes are taken in the coordinate-independent form and
coupling to higher modes is treated as a perturbation. Here, the boundary
condition in Eq. (\ref{boundary}) requires the use of rotation to a local
spin-coordinate system, $\hat{P}=\hat{O}\widehat{\tilde{P}}\hat{O}^{-1}$,
prior to making the 0D approximation, with $\hat{O}=\exp \{i[\hat{S}%
_{1}X_{2}\lambda _{2}^{-1}-\hat{S}_{2}X_{1}\lambda _{1}^{-1}]\}\exp
\{-i\varphi _{s}(\vec{X})\hat{S}_{3}\}\exp \{-i\varphi _{A}(\vec{X})\}$
where harmonic functions $\varphi $ transform the symmetric gauge in Eq. (%
\ref{aperp}) to such a gauge, where vector potentials on the boundary are
tangential to it. This eliminates the lowest orders SO coupling terms from
the boundary condition, and, in a small dot \cite{Largesample} $L_{1,2}\ll
\lambda _{1,2}$, can be followed by a perturbative analysis of extra terms
generated by rotation $\hat{O}$ in Eqs. (\ref{Coop}). This step results in
the 0D matrix equation for the Cooperon, 
\end{multicols}
\widetext
\begin{equation}
\hat{P}=\left[ \gamma \hat{\delta}+i\epsilon _{{\rm Z}}\hat{\eta}+\left( 
\hat{\delta}\sqrt{\tau _{B}^{-1}}-\sqrt{\epsilon _{\bot }^{{\rm so}}}\hat{S}%
_{3}\right) ^{2}+\epsilon _{\bot }^{{\rm Z}}(\hat{\delta}-\hat{S}%
_{3}^{2})+\epsilon _{\Vert }^{{\rm so}}(\widehat{\vec{S}^{2}}-\hat{S}%
_{3}^{2})\right] ^{-1}. 
\label{CoopMatrix}
\end{equation}

\begin{multicols}{2}

The form of Eq. (\ref{CoopMatrix}) is applicable beyond the diffusive
approximation as it follows from purely the symmetry considerations. The
difference in the third term in brackets reflects the addition or
subtraction of the Berry and Aharonov-Bohm phases, as was pointed out in
Ref. \cite{Aronov}. The expression for the weak localization correction can
be found from Eq. (\ref{CoopMatrix}) as $g_{{\rm wl}}\propto {\rm tr}\{\hat{P%
}[\hat{\delta}-\widehat{\vec{S}^{2}}]\}$. In a dot with $\tau _{B}^{-1}=0$
and $\gamma ,\,\epsilon _{{\rm Z}},D/\lambda _{1,2}^{2}\ll E_{{\rm T}}$ \cite
{Largesample}, this yields: 
\begin{equation}
\frac{4g_{{\rm wl}}}{\xi }\approx -\frac{\gamma }{\gamma +\epsilon _{\bot }^{%
{\rm so}}}-\frac{\gamma }{\gamma +\epsilon _{\bot }^{{\rm Z}}+2\epsilon
_{\Vert }^{{\rm so}}}+\frac{\epsilon _{\bot }^{{\rm so}}}{\gamma +\epsilon
_{\bot }^{{\rm so}}+\frac{\epsilon _{{\rm Z}}^{2}}{\gamma }},
\label{crossoverWL}
\end{equation}
where we used the fact that $\epsilon _{\bot }^{{\rm Z}}\ll \epsilon _{{\rm Z%
}}$. It is interesting to notice that the application of the Zeeman field
alone (in plane magnetic field\cite{MarcusSpin}) does not suppress the weak
localization completely at $\epsilon _{{\rm Z}}\ll E_{\text{T}}$. In the
opposite case of $\epsilon _{{\rm Z}}\gtrsim E_{\text{T}}$, it does 
(though has to be studied beyond the universal limit). However, the effect  
of such a strong in-plane fields on orbital motion becomes already sufficient 
to suppress the weak localization \cite{Jung}.

The form of Eq. (\ref{crossoverWL}) in the experimentally easier achievable
'high'-energy crossover 1$\rightarrow $4$\rightarrow $7\ is limited by only
two first terms and suggests a possible procedure for measuring the ratio $%
\lambda _{1}/\lambda _{2}$. By fitting experimental magnetoresistance data
to $g_{{\rm wl}}(\epsilon _{{\rm Z}})$ in Eq. (\ref{crossoverWL}), one would
determine the characteristic in-plane field ${\cal B}$. For a dot with a
strongly anisotropic shape, such a parameter would depend on the orientation
of an in-plane magnetic field. In particular, ${\cal B}$ can be measured for
two orientations of $\vec{B}=B\vec{l}$: namely, ${\cal B}_{[110]}$ for $\vec{%
l}=[110]$ and ${\cal B}_{[1\bar{1}0]}$ for $\vec{l}=[1\bar{1}0]$. One should
also make a simultaneous measurement of two characteristic fields ${\cal B}%
_{[110]}^{\prime }$ and ${\cal B}_{[1\bar{1}0]}^{\prime }$ in a dot produced
on the same chip by rotating the same lithographic mask by 90${{}^\circ}$. 
The anisotropy of the SO coupling can be then obtained directly from the
ratio 
\[
\left( {\cal B}_{[110]}{\cal B}_{[110]}^{\prime }/{\cal B}_{[1\bar{1}0]}%
{\cal B}_{[1\bar{1}0]}^{\prime }\right) =\left( \lambda _{1}/\lambda
_{2}\right) ^{4}, 
\]
independently of the details of sample geometry.

The other interesting feature may be observed in the weak localisation $g_{%
{\rm wl}}$ in the regime of a crossover 2$\rightarrow $2u driven by a weak
perpendicular magnetic field for $\epsilon _{{\rm Z}}=0$. Since the SO
coupling effect in Eqs. (\ref{aperp},\ref{CoopMatrix}) acts as a homogeneous
magnetic field distinguishing between up- and down-spin electrons, the
external field can be used to compensate the effect of the SO coupling for
one spin component \cite{Aronov}, which would produce a dip in the weak
localisation correction. Indeed, at $\gamma \ll \epsilon _{\bot }^{{\rm so}}$, the
function 
\begin{eqnarray}
\frac{2g_{{\rm wl}}}{\xi } &\approx &\frac{\gamma \epsilon _{\Vert }^{{\rm so%
}}}{\left( \gamma +\tau _{B}^{-1}\right) \left( \gamma +\tau
_{B}^{-1}+2\epsilon _{\Vert }^{{\rm so}}\right) } \\
&&-\frac{\gamma \left( \gamma +\tau _{B}^{-1}+\epsilon _{\bot }^{{\rm so}%
}+\epsilon _{\Vert }^{{\rm so}}\right) }{\left( \gamma +\tau
_{B}^{-1}+\epsilon _{\bot }^{{\rm so}}+\epsilon _{\Vert }^{{\rm so}}\right)
^{2}-4\tau _{B}^{-1}\epsilon _{\bot }^{{\rm so}}},  \nonumber
\end{eqnarray}
has a minimum at the value of the field $2e{\cal B}_{z}=c\hbar /(\lambda_{1}\lambda _{2})$,
independently of the sample geometry, which should provide a very accurate measurement
of  $\lambda_{1}\lambda _{2}$.

We thank P.W. Brouwer, and C.M. Marcus for discussions. This research was
funded by EPSRC, Packard Foundation and NATO CLG.

\end{multicols}


\begin{references}
\bibitem{Remark-perpfield}  Zeeman field is taken parallel to the plane of
2D gas, $\vec{l}=(l_{x},l_{y})$. The effect of its normal component is
negligible when $B_{z}{\cal A}<\Phi _{0}$, since $\epsilon _{{\rm Z}}(\Phi
_{0}/{\cal A})\sim (g^{\ast }m/m_{e})\Delta \ll \Delta $.
\bibitem{Largesample}  If $\lambda _{1,2}\ll L_{1,2}$, the use of a gauge
transformation is not helpful, and the spin relaxation problem has to be
treated as a 2D one. This reproduces the Dykonov-Perel relaxation \cite
{DykonovPerel} with the rates $\epsilon _{{\rm DP}}^{{\rm so}}\sim D/\lambda
^{2}\gg E_{{\rm T}}$, associated with the weak localisation scenario, where
the symplectic ensemble at $\epsilon _{{\rm Z}}=0$ transforms into the
unitary ensemble at $\epsilon _{{\rm Z}}^{2}/\epsilon _{{\rm DP}}^{{\rm so}%
}>\tau _{{\rm esc}}^{-1}$.
\end{references}
\end{document}